\documentclass[conference]{IEEEtran}

\usepackage{times}
\usepackage{amsmath}
\usepackage{amsfonts}
\usepackage{amssymb}
\usepackage{amstext}
\usepackage{latexsym}
\usepackage{color}
\usepackage{ifthen}
\usepackage{multirow}
\usepackage{verbatim}
\usepackage{array,tabularx}

\usepackage[pdftex]{graphicx}

\newcommand{\be}{\begin{equation}}
\newcommand{\ee}{\end{equation}}
\newcommand{\bear}{\begin{eqnarray}}
\newcommand{\eear}{\end{eqnarray}}
\newcommand{\bears}{\begin{eqnarray*}}
\newcommand{\eears}{\end{eqnarray*}}
\newcommand{\bi}{\begin{itemize}}
\newcommand{\ei}{\end{itemize}}
\newcommand{\ben}{\begin{enumerate}}
\newcommand{\een}{\end{enumerate}}

\newtheorem{theorem}{Theorem}

\newtheorem{example}[theorem]{Example}
\newtheorem{corollary}[theorem]{Corollary}

\IEEEoverridecommandlockouts

\renewcommand{\baselinestretch}{1}

\begin{document}

\title{Capacity and Security of Heterogeneous Distributed Storage Systems}
\author{
\IEEEauthorblockN{Toni Ernvall, Salim El Rouayheb, \emph{Member, IEEE},\\ Camilla Hollanti, \emph{Member, IEEE}, and H. Vincent Poor, \emph{Fellow, IEEE} }
\thanks{T. Ernvall is with the  Department of Mathematics, University of Turku, Finland,
e-mail: {tmernv}@utu.fi.}
\thanks{S. El Rouayheb and H. V. Poor   are with the Department of Electrical Engineering, Princeton University, e-mails: {salim, poor}@princeton.edu.}
        \thanks{C. Hollanti is with the Department of Mathematics and System Analysis,  Aalto University,
        Finland,  e-mail: {camilla.hollanti}@aalto.fi.}
\thanks{This research was supported by the U.\ S. National Science Foundation under Grant CCF-1016671 and by the Academy of Finland under Grant \#131745.}

}
\maketitle

\begin{abstract}
We study the capacity of \emph{heterogeneous} distributed storage systems under repair dynamics. Examples of these systems include peer-to-peer storage clouds, wireless, and Internet caching systems. Nodes in  a heterogeneous system  can have different storage capacities and different repair bandwidths.   We give lower and upper bounds on the system capacity. These bounds  depend on either the average resources per node, or on a detailed knowledge of the node characteristics. Moreover, we study the case in which  nodes  may be compromised by an eavesdropper, and give bounds on the system secrecy capacity. One implication of our results
is that symmetric repair maximizes the capacity of a homogeneous system, which justifies the model widely used in the literature.

\end{abstract}

\section{Introduction}\label{sec:Intro}

Cloud storage  has emerged in recent years as an inexpensive and scalable solution for storing large amounts of data and making it pervasively available to  users. The growing success of cloud storage has been accompanied by new advances in the theory of such systems, namely the application of network coding techniques for distributed data storage and the theory of regenerating codes introduced by Dimakis \emph{et al.} \cite{DGWWR07}, followed by a large body of further work in the literature.

Cloud storage systems are typically built using a large number of inexpensive commodity disks that fail frequently, making failures ``the norm rather then the exception" \cite{GFS}. Therefore, it is a prime concern to achieve fault-tolerance in these systems and minimize the probability of losing the stored data.
The recent theoretical results uncovered fundamental tradeoffs among system resources (storage capacity, repair bandwidth, etc.) that are necessary to achieve fault-tolerance. They also provided  novel  code constructions  for data redundancy schemes that can achieve  these tradeoffs in certain cases; see for example \cite{DRWS10, RSK10} and \cite{RR10}.

The majority of the results in the literature of this field focus on a homogeneous model  when studying the information theoretic limits  on the performance of  distributed storage systems. In a homogeneous system all the  nodes (hard disks or other storage devices) have the same parameters (storage capacity, repair bandwidth, etc.). This model  encompasses many real-world storage systems such as clusters in a data center, and has been instrumental in forming the engineering intuition for understanding these systems.
Recent development have included the emergence  of  \emph{heterogeneous} systems that pool together  nodes from different sources and with different characteristics to form one big reliable cloud storage system. Examples   include peer-to-peer (p2p), or hybrid (p2p-assisted) cloud storage systems   \cite{Oceanstore, SpaceMonkey}, Internet caching systems for video-on-demand applications \cite{ZCPK11, PRZKR11}, and caching systems in heterogeneous wireless networks \cite{GDM12}. Motivated by these applications, we  study  the capacity of heterogeneous distributed storage systems (DSS) here under reliability and secrecy constraints.

\noindent\paragraph*{Contributions}

The capacity of a DSS is defined as the maximum amount of information that can be delivered to any user contacting $k$ out of $n$ nodes in the system. Intuitively, in a heterogeneous system,  this capacity should be limited by the ``weakest" nodes. However, nodes can have different storage capacities and different repair bandwidths. And the tension between these two set of parameters makes it challenging  to identify which nodes are the ``weakest".

  Our first result  establishes an upper bound on the capacity of a DSS  that depends on  the average resources in the system (average storage capacity  and average repair bandwidth per node). We use this bound to prove that symmetric repair, \emph{i.e.}, downloading equal amount of data from each helper node, maximizes the capacity of a homogeneous DSS. While the optimality of symmetric repair is known for the special case of MDS codes \cite{Yunnan}, our results assert  that symmetric repair is always optimal for any choice of system parameters.
   Further, our proof avoids the combinatorial cut-based arguments typically used this context.

  In addition, we give an expression for the capacity when we know the characteristics of all the nodes in the system (not just the averages). This expression may be hard to compute,  but  we use it to derive additional bounds that are easy to evaluate.
   Our techniques generalize to the  scenario in which  the system is compromised by an eavesdropper\footnote{Our results also generalize to the case of a malicious adversary who can corrupt the stored data. This model will be included in the extended version of this paper.}. We give bounds on the secrecy capacity when the system is supposed to leak no information to the eavesdropper (perfect secrecy).
Here too, we show that symmetric repair maximizes the secrecy capacity of a homogeneous system.

\noindent\paragraph*{Related work} Wu proved the optimality of symmetric repair  in \cite{Yunnan} for the special case of a DSS using Maximum Distance Separable (MDS) codes. Coding schemes for a non-homogeneous storage system with one super-node that is more reliable and has more storage capacity were studied in \cite{VYL12}. References \cite{LDH12} and \cite{VGA12} studied the problem of storage allocations in distributed systems under a total storage budget constraint. Pawar \emph{et al.}\ \cite{PRK10, PRK11} studied the secure capacity  of distributed storage systems under eavesdropping and malicious attacks.

\noindent\paragraph*{Organization} Our paper is organized as follows. In Section~\ref{sec:Model}, we describe our model for heterogeneous DSS and set up the notation. In Section~\ref{sec:results}, we summarize our main results. In Section~\ref{sec:Cap}, we prove our bounds on the capacity of a heterogeneous DSS. In Section~\ref{sec:security}, we study the secrecy capacity in the presence of an eavesdropper. We conclude in Section~\ref{sec:conclusion} and discuss some open problems. We postpone some of the  proofs  to the Appendix, where we also discuss  the generalizability  of our results from  functional to exact repair.

\section{Model}\label{sec:Model}
A heterogeneous distributed storage system  is formed of $n$ storage nodes  $v_1,\dots, v_n$ with storage capacities  $\alpha_1, \dots, \alpha_n$ respectively. Unless stated otherwise, we assume that the nodes are indexed in increasing order of capacity, {\em i.e.}, $\alpha_1 \leq \alpha_2 \leq \dots \leq \alpha_n$.  In a homogeneous system all nodes have the same storage capacity $\alpha$, \emph{i.e.}, $\alpha_i=\alpha, \forall i$. As a reliability requirement, a user should be able to obtain a file by contacting any $k<n$ nodes in the DSS. The nodes forming the system are unreliable and can fail.  The system is \emph{repaired} from a failure by replacing the failed node with a new node. Upon joining the system, the new node downloads its data from $d$, $k\leq d\leq n-1$, helper nodes in the system.

The repair process can be either \emph{exact} or \emph{functional}. In the case of exact repair, the new node is required to store an exact copy of the data that was stored on the failed node. Whereas in the case of  functional repair, the data stored on the new node does not have to be an exact copy of the lost data, but  merely ``functionally equivalent" in the sense that it preserves the property that contacting any $k$ out of $n$ nodes is sufficient to reconstruct a stored file. We focus on functional repair in this paper, although some of our results do generalize to the exact repair model (see the discussion in Appendix~\ref{AppendixZ}).

An important system parameter is  the $\emph{repair bandwidth}$ which refers to the total amount of data downloaded by the new node. In a homogeneous system, the repair bandwidth, denoted by $\gamma$,  is the same for any new node joining the system.  The typical model adopted in the literature  assumes    \emph{symmetric repair} in which the total repair bandwidth $\gamma$ is divided equally among the $d$ helpers. Thus, the new node downloads $\beta=\gamma/d$  amount of information from each helper.
In a heterogeneous system the repair bandwidth can vary depending on  which node has failed and which nodes are helping in the repair process. We denote by  $\beta_{ijS}$ the amount of information that a new node replacing the failed node $v_j$ is downloading from   helper node $v_i$ when the other helper node belong to the index set $S$  ($i\in S, |S|=d$). An important special case is when the repair bandwidth per helper  depends only on the identity of the helper node and not on the identity of the failed node or  the other helpers. In this case, we say that helper node $v_i$ has repair bandwidth $\beta_{i}$, \emph{i.e.}, $\beta_{ijS}=\beta_i,  \forall j,S$. In the case of a homogeneous system with symmetric repair, we have $\beta_{ijS}=\beta=\gamma/d,  \forall i, j,S$.

We focus on repair from single node failures\footnote{Multiple failures can be repaired independently as long as there are at least $d$ helper nodes in the system. For another model of repair that assumes cooperation when repairing multiple failures in homogeneous systems, refer to \cite{SH12} and \cite{KSS11}.}. In this case, there are $\binom{n-1}{d}$ possibilities for the set of helpers $S$. Therefore,  the average repair bandwidth $\gamma_{j}$ of node $v_j$ is
\begin{equation}\label{eq:AvBW}
\gamma_{j} = \binom{n-1}{d}^{-1} \sum_{\substack{
   S:  j \notin S \\
   |S|=d
  }} \sum_{i \in S} \beta_{ijS}.
\end{equation}
We denote by $\overline{\gamma} = \frac{1}{n}\sum_{j=1}^{n} \gamma_j$  and $\overline{\alpha} = \frac{1}{n}\sum_{j=1}^{n} \alpha_j$ the average total repair bandwidth and  average node capacity in the DSS, respectively.

We are interested in finding the capacity $C$  of a heterogeneous system. The capacity $C$ represents  the maximum amount of information that can be downloaded by any user contacting  $k$  out of the $n$ nodes in the system. Recall from \cite{DGWWR07}, that the capacity $C^{ho}$ of a homogeneous system implementing symmetric repair is given by
\begin{equation}\label{eq:AlexCap}
C^{ho}(\alpha,\gamma)=\sum_{i=1}^{k}\min\left\{\alpha,(d-i+1)\frac{\gamma}{d}\right\}.
\end{equation}

We are also interested in characterizing the secrecy capacity of the system when some nodes are compromised by an eavesdropper. We follow the model in \cite{PRK10} and \cite{PRK11} and denote by $\ell$, $\ell\leq k$,  the number of compromised nodes. The eavesdropper is assumed to be passive. She can read the data downloaded during repair and  stored on a compromised node. We are interested here in information theoretic secrecy which characterizes the fundamental ability of the system to provide data confidentiality independently of cryptographic methods. The secrecy capacity of the system, denoted by $C_s$, is defined as the maximum amount of information that can be delivered to a user  without revealing any information to the eavesdropper (perfect secrecy).  We denote by $C_s^{ho}$ the secrecy capacity of a  homogeneous system with symmetric repair. Finding $C_s^{ho}$ is still an open problem in general. The following upper bound was shown to hold  in \cite{PRK10} and \cite{PRK11}:
\begin{equation}\label{eq:secCap}
C_s^{ho}(\alpha,\gamma,\ell)\leq\sum_{i=\ell+1}^{k}\min\left\{\alpha,(d-i+1)\frac{\gamma}{d}\right\}.
\end{equation}

\begin{figure*}[t]
  \begin{center}
   \includegraphics[]{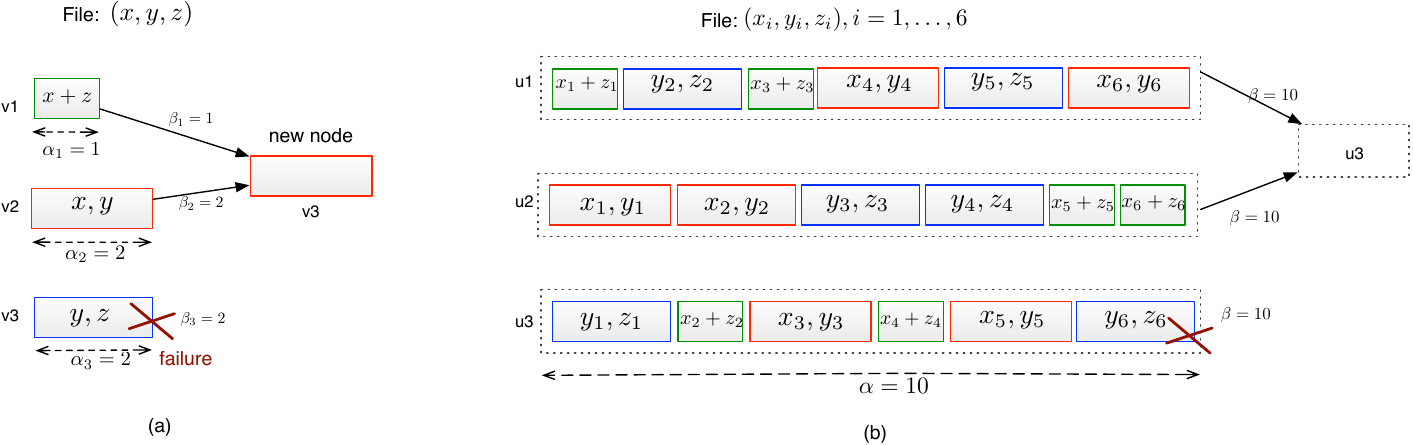}
\caption{ An example that illustrates the proof of the upper bound \eqref{eq:HCap}  on the capacity of a heterogeneous system. (a)  A heterogeneous distributed storage system (DSS) with $(n,k,d)=(3,2,2)$. The nodes have storage capacities $\alpha_1=1,\alpha_2=\alpha_3=2$ and the repair bandwidth per helper are  $\beta_1=1, \beta_2= \beta_3=2$. (b) A DSS constructed by combining together $n!=6$ copies of the original heterogeneous system corresponding to all possible node permutations. The obtained DSS is homogeneous with uniform storage per node $\alpha=10$ and repair bandwidth per helper $\beta=10$.  The capacity of this system is 20 as given by \eqref{eq:AlexCap} \cite{DGWWR07}. Any code that stores a file of size $C$ ($C=3$ here) on the original DSS can be transformed into a scheme that stores a file of size $n!C=6C$ in the ``bigger" system. This gives the upper bound in \eqref{eq:HCap} $C\leq20/6=10/3$.
}
    \label{fig:DSS}
   \end{center}
\end{figure*}

\section{Main Results}\label{sec:results}
We start by summarizing our results.  Theorem~\ref{th:HCap} gives a general upper bound on the storage capacity of a heterogeneous DSS as a function of the  average resources per node.

\begin{theorem}\label{th:HCap}
The capacity $C$ of a heterogeneous distributed storage system, with node average capacity $\bar{\alpha}$ and average  repair bandwidth  $\bar{\gamma}$, is upper bounded by
\begin{equation}\label{eq:HCap}
C\leq \sum_{i=1}^{k}\min\left\{\bar{\alpha},(d-i+1)\frac{\bar{\gamma}}{d}\right\}=C^{ho}(\bar{\alpha},\bar{\gamma}).
\end{equation}
\end{theorem}

The right-hand side term in \eqref{eq:HCap} is the capacity of a homogeneous system in  \eqref{eq:AlexCap} in which all nodes have storage $\alpha=\bar{\alpha}$ and total repair bandwidth $\gamma=\bar{\gamma}$ . Th.~\ref{th:HCap} states  that the capacity of a DSS cannot exceed that of a homogeneous system where the total system resources are split equally among all the nodes.  Moreover, Th.~\ref{th:HCap} implies that \emph{symmetric repair is optimal} in homogeneous systems in the sense that it maximizes the system capacity. This justifies the repair model adopted in the literature.
This result is stated formally in Cor.~\ref{cor:Sym}.

While the optimality of symmetric repair is known for the special case of MDS codes \cite{Yunnan}, Cor.~\ref{cor:Sym} asserts  that symmetric repair is always optimal for any choice of system parameters.  This result  follows directly from Th.~1 and  avoids the combinatorial cut-based arguments that may be needed in a more direct proof.

\begin{corollary}\label{cor:Sym}
In a homogeneous DSS with node capacity $\alpha$ and total repair bandwidth $\gamma$, symmetric repair maximizes the system capacity.
\end{corollary}

When we know the parameters of the nodes in the system beyond the averages, we can obtain possibly tighter bounds as described in Th.~\ref{th:upnlw}. To simplify the notation, let us order the repair bandwidth per helper  $\beta_{ijS}$ into an increasing sequence  $\beta_{1}',\beta_{2}',\dots,\beta_{m}'$, such that  $\beta_l' \leq \beta_{l+1}'$  and where $m=nd\binom{n-1}{d}$. Also, recall that $\alpha_1 \leq \alpha_2 \leq \dots \leq \alpha_n$.

\begin{theorem}\label{th:upnlw}
The capacity $C$ of heterogeneous DSS is bounded by
$$
C_{\min} \leq C \leq C_{\max}
$$
where
$$
C_{\min} = \min_{l=0,\dots,k} \left( \sum_{j=1}^{l} \alpha_j + \sum_{j=1}^{h} \beta_{j}' \right),
$$
$$
C_{\max} = \min_{l=0,\dots,k} \left( \sum_{j=1}^{l} \alpha_j + \sum_{j=1}^{h} \beta_{m+1-j}' \right),
$$
and
$$
h=\frac{(2d-k-l+1)(k-l)}{2}.
$$
\end{theorem}

When the system is compromised by an eavesdropper the system secrecy capacity can be upper bounded as follows.
\begin{theorem}\label{th:upsec}
The secrecy capacity $C_s$ of a DSS when $\ell$ nodes in the system are compromised by an eavesdropper is upper bounded by
\begin{equation}\label{eq:sechet}
C_s\leq \sum_{i=\ell+1}^{k} \min\left\{ \overline{\alpha}, (d-i+1)\frac{\overline{\gamma}}{d} \right\} .
\end{equation}

\end{theorem}
This theorem implies that symmetric repair also maximizes the secrecy capacity of a homogeneous DSS.

\section{Capacity of Heterogeneous DSS}\label{sec:Cap}

\subsection{Example \& Proof of Theorem~\ref{th:HCap}}\label{sec:ToyEx}

We illustrate the proof of Th.~\ref{th:HCap} through an example for  the special case in which the  bandwidths depend only on identity of  the helper node. We compute the capacity of the DSS for this specific example, and show that it is strictly less than the upper bound of Th.~\ref{th:HCap}. That is, it does not achieve the capacity of a homogenous system with the same average characteristics. More specifically, consider the heterogeneous DSS depicted in Fig.~\ref{fig:DSS}(a) with $(n,k,d)=(3,2,2)$ formed of $3$ storage nodes $v_1,v_2$ and $v_3$  with storage capacities  $(\alpha_1,\alpha_2,\alpha_3)=(1,2,2)$ and repair bandwidths $(\beta_1,\beta_2,\beta_3)=(1,2,2)$. The average node capacity $\overline{\alpha}=5/3$ and repair bandwidth  are  $\overline{\beta}=10/3$. Th.~\ref{th:HCap} gives that the capacity of this DSS $C\leq 10/3=3.33$.

For this example, it is easy to see that the DSS capacity is  $C=3\leq 10/3$. In fact, a user contacting nodes $v_1$ and $v_2$ cannot download more information then their total storage $\alpha_1+\alpha_2=3$. This upper bound is achieved by the code in {Fig.~\ref{fig:DSS}(a)}. The code stores a file of 3 units $(x,y,z)$ in the system. During repair the new node downloads the whole file and stores the lost piece of the data (note that the repair bandwidth constraints allow this trivial repair).

To obtain the  upper bound in \eqref{eq:HCap}, we use the original heterogeneous DSS  to construct a  ``bigger"  homogeneous system. We obtain this new system by ``glueing"  together  $n!=3!=6$ copies of the original  DSS as shown in Fig.~\ref{th:HCap}(b). Each copy corresponds to a different permutation of the nodes. In the figure,  the $i^{\rm th}$ copy stores the file $(x_i,y_i,z_i)$. For example in Fig.~\ref{th:HCap}(b), the first copy is the original system itself, the second corresponds to node $v_1$ and node $v_3$ switching positions, and so on.

The ``bigger" system is homogeneous because all  its nodes have storage $\alpha=10$ and repair bandwidth per helper $\beta=\gamma/d=10$. The capacity $C'$ of this system can be computed from \eqref{eq:AlexCap}:
\begin{equation}\label{eq:HomCap}
\mathcal{C'} = \sum_{i=1}^{k} \min \{ \alpha , (d-i+1)\frac{\gamma}{d} \}=20.
\end{equation}

As seen in Fig.~\ref{fig:DSS}, any scheme that can store a file of size $C$ in the original DSS can be transformed into a scheme that can store a file of size $n!C$ in the ``bigger" DSS. Therefore, we get $n!C\leq C'$ and $C\leq=10/3$. This argument can be directly generalized to arbitrary heterogeneous systems.
The general proof  follows the same steps explained above and can be found in Appendix~\ref{AppendixA}.

Theorem~\ref{th:HCap} implies that symmetric repair, {\em i.e.}, downloading equal numbers of bits from each of the helpers, is optimal in a homogeneous system. To see this, consider a DSS with node storage capacity $\alpha$, and a total repair bandwidth budget $\gamma$. A new node joining the system has the flexibility to  arbitrarily split its repair bandwidth among the $d$ helpers as long as the total amount of downloaded information does not exceed $\gamma$. In other words, we have $\sum_{i\in S }\beta_{ijS}=\gamma, \forall j,S.$ Now, irrespective of how each new node splits its   bandwidth budget, the average repair bandwidth in the system is the same, $\overline{\gamma}=\gamma$. If we apply Th.~\ref{th:HCap},  we get an upper bound that matches exactly the capacity in \eqref{eq:AlexCap} of a homogeneous DSS with symmetric repair. Hence, we obtain the result in Cor.~\ref{cor:Sym}.

\subsection{Proof of Theorem~\ref{th:upnlw}}

To avoid heavy notation, we focus on  the case in which the repair bandwidth depends only on the helper node ($\beta_{ijS}=\beta_i$).   We give in Th.~\ref{th:upnlw2} lower and upper bounds  specific to this case. These    bounds are similar to the ones in Th.~\ref{th:upnlw}, but can be tighter. The proof of Th.~\ref{th:upnlw} follows the exact steps of the proof below and will be omitted here. Again, we assume that the nodes are indexed in increasing order of node capacity, $\alpha_1\leq\alpha_2\leq\dots\leq\alpha_n$. We also order the values of the  repair bandwidths $\beta$  to obtain the increasing sequence $\beta_1'\leq\beta_2'\leq\dots\leq\beta_n'$.
\begin{theorem}\label{th:upnlw2}
The capacity $C$ of a heterogeneous DSS, in which the repair bandwidth depends only on  the identity of the helper node, is bounded as  $C_{\min}' \leq C \leq C_{\max}'$,
where
\begin{equation}\label{eq:Cpmin}
\begin{split}
C_{\min}' & = \sum_{i=1}^{k} \min(\alpha_i , \beta_1' + \beta_2' + \dots + \beta_{d-i+1}') \\
          & = \min_{l=0,\dots,k} \left( \sum_{i=1}^{l} \alpha_i + \sum_{j=0}^{k-l-1}  \sum_{i=1}^{d-l-j} \beta_i' \right),
\end{split}
\end{equation}
and
\begin{equation}\label{eq:Cpmax}
\begin{split}
C_{\max}' & = \sum_{i=1}^{k} \min(\alpha_i , \beta_{i+1}' + \beta_{i+2}' + \dots + \beta_{d+1}') \\
          & = \min_{l=0,\dots,k} \left( \sum_{i=1}^{l} \alpha_i + \sum_{j=1}^{k-l}  \sum_{i=l+1+j}^{d+1} \beta_i' \right).
\end{split}
\end{equation}
\end{theorem}

The second expressions for $C_{\min}'$ and $C_{\max}'$ highlight the analogy with the bounds in Th.~\ref{th:upnlw}. Before proving Th.~\ref{th:upnlw2}, we give a couple of  illustrative examples and discuss some special cases.
\begin{example}
Consider again the example in the previous section where $(n,k,d)=(3,2,2)$ and where the nodes parameters are $(\alpha_1,\beta_1)=(1,1),$ $(\alpha_2,\beta_2)=(\alpha_3,\beta_3)=(2,2)$. Here, $C_{\min}'=2$ and $C_{\max}'=3$.  Note that here $C_{\max}'$ is tighter then the average-based upper  bound of Th.~\ref{th:HCap} which gives $C\leq3.33$. Recall that the capacity for this system is $C=3=C_{\max}'$.
\end{example}

\begin{example}
Consider now a second DSS with $(n,k,d)=(3,2,2)$ and  $(\alpha_1,\beta_1)=(5,3),(\alpha_2,\beta_2)=(6,4)$ and $(\alpha_3,\beta_3)=(7,5)$. Here, $C_{\min}'=9$ and $C_{\max}'=11$, and  Th.~\ref{th:HCap} gives $C\leq10<C_{\max}'$.\end{example}

The upper and lower bounds can coincide ($C_{\min}'=C_{\max}'$) in certain cases, which gives the exact expression of the capacity. For example:
\begin{enumerate}
\item A homogeneous DSS, where we recover the capacity expression in \eqref{eq:AlexCap}.
\item A DSS with uniform repair bandwidth, \emph{i.e.}, $\beta_i=\beta, \forall i$. The capacity is
$C= \sum_{i=1}^{k} \min(\alpha_i , (d-i+1)\beta)$.
\item Whenever $\alpha_i\leq\beta_1', \forall i$. In this case the  capacity $ C= \sum_{i=1}^{k} \alpha_i$.
\end{enumerate}

\begin{figure}[t]
  \begin{center}
   \includegraphics[]{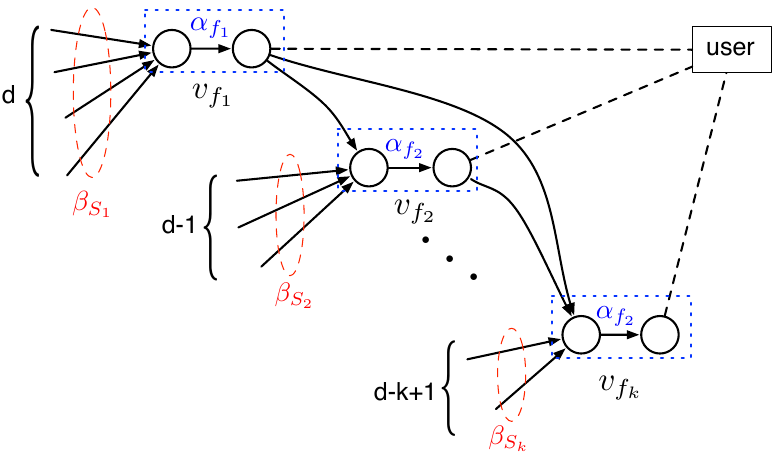}
\caption{ A series of $k$ failures  and repairs in the DSS that explains the capacity expression in \eqref{eq:exactCap}. Nodes $v_{f_1},\dots,v_{f_k}$ fail successively and are repaired as depicted above. The amount of ``new" information that node $v_{f_i}$ can give the user is the minimum between his storage capacity $\alpha_{f_i}$ and downloaded data $\beta_{S_i}$.  }
    \label{fig:cut}
   \end{center}
\end{figure}
To prove the upper and lower bounds in Th.~\ref{th:upnlw2}, we first establish the following expression of the DSS capacity.
\begin{theorem}\label{th:cap}
The capacity $C$ of  a heterogeneous DSS is given by
\begin{equation}\label{eq:exactCap}
C = \min_{\substack{
(f_1, \dots, f_k) \\
f_i \neq f_j \text{ for } i \neq j }} \sum_{i=1}^{k} \min \left( \alpha_{f_i} , \min_{\substack{
|S_i|=d+1-i \\
 S_i\cap \{ f_1 , \dots, f_i \} = \emptyset }} \beta_{S_i} \right),
\end{equation}
where for any  $ S\subseteq\{1,\dots,n\}, \beta_{S} = \sum_{i \in S} \beta_i.$
\end{theorem}

The proof of  Th.~\ref{th:cap} is a generalization of the proof in   \cite{DGWWR07} of the capacity of a homogeneous system \eqref{eq:AlexCap}. We defer this proof to Appendix~\ref{sec:A2} and  explain here the intuition behind it.
Consider the scenario depicted in Fig.~\ref{fig:cut} where nodes $v_{f_1},\dots,v_{f_k}$ fail and are repaired successively such that node $v_{f_i}$ is repaired by downloading data from the previously repaired  nodes $v_{f_1},\dots,v_{f_{i-1}}$ and $d-(i-1)$ other helper nodes in the system. Consider now a user contacting nodes $v_{f_1},\dots,v_{f_k}$.

The amount of ``non-redundant" information that node $v_{f_i}$ can give to the user is evidently limited by its storage capacity $\alpha_i$ on one hand,   and on the other hand, by the amount of information $\beta_{S_i}$ downloaded from the  $d-i+1$ helper nodes that are not connected to the user. Minimizing over all the choices of $f_1,\dots, f_k$ gives the expression in \eqref{eq:exactCap}.

It is not clear whether the capacity expression in \eqref{eq:exactCap} can be computed efficiently. For this reason we give upper and lower bounds that are easy to compute. To get the lower bound in \eqref{eq:Cpmin},  let $(f_1, \dots, f_k)=(f_1^*, \dots, f_k^*)$ be the minimizer of \eqref{eq:exactCap}.  We have
\begin{equation}
\begin{split}
C
&= \sum_{i=1}^{k} \min \left( \alpha_{f_i^*} , \min_{\substack{
|S_i|=d+1-i \\
\{ f_1^* , \dots, f_i^* \} \cap S_i = \emptyset }} \beta_{S_i} \right) \\
& \geq \sum_{i=1}^{k} \min \left( \alpha_{f_i^*} , \beta_1' + \beta_2' + \dots + \beta_{d-i+1}' \right) \\
& \geq \sum_{i=1}^{l^*} \alpha_{i} + \sum_{i=1}^{d-l^*} \beta_{i}' + \sum_{i=1}^{d-l^*-1} \beta_{i}' + \dots + \sum_{i=1}^{d-k+1} \beta_{i}' \\
&= \min_{l=0,\dots,k} \left( \sum_{i=1}^{l} \alpha_i + \sum_{j=0}^{k-l-1}  \sum_{i=1}^{d-l-j} \beta_i'\right),
\end{split}
\end{equation}
where  $l^*, 0\leq l^*\leq k$ is the number of those cases where $\alpha_{f_i^*}$ is smaller or equal than the corresponding sum of $\beta$'s.

The upper bound $C_{\max}' $ is obtained by taking $(f_1, \dots, f_k)=(1, \dots, k)$ in \eqref{eq:exactCap} and following similar steps as above.

\section{Security}\label{sec:security}
\subsection{Secrecy Capacity}
We now consider the case in which $\ell$ nodes in the system are compromised by a passive eavesdropper who can observe their downloaded and stored data, but cannot alter it. The \emph{secrecy capacity}  $C_s$ of the system is the maximum amount of information that can be delivered to any user  without revealing any information to the eavesdropper (perfect secrecy).

Formally, let $S$ be the information source that represents the file that is stored on the DSS.  A user contacts the nodes in any set $B\subset\{v_1,\dots,v_n\}$ of size $k$ and downloads their stored data denoted by $C_B$. The user should be able to decode the file, which implies $H(S|C_B)=0$. Let $E$ be the set of the $\ell$ compromised nodes, and   $D_E$ be the data observed by the eavesdropper. The perfect secrecy condition implies that   $H(S|D_E)=H(S)$. Following the definition in \cite{PRK11}, we write the secrecy capacity as\begin{equation}
C_s(\alpha,\gamma) = \sup_{\substack{
           H(S|C_B)=0 \forall B \\
           H(S|D_E)=H(S) \forall E}} H(S).
\end{equation}

Finding the secrecy capacity of a DSS is a hard problem and is still open in general, even for the class of  homogeneous systems. Let $C_s^{ho}(\alpha,\beta,\ell)$ denote  the secrecy capacity of a homogeneous DSS implementing symmetric repair  and having $\ell$ compromised nodes. Following the same steps in the proof of Th.~\ref{th:HCap}, we can show that the secrecy capacity $C_s$ of a heterogeneous DSS cannot exceed that of a homogeneous DSS having the same average resources.

\begin{theorem}\label{eq:secgenup}
Consider a heterogeneous DSS with average  storage capacity per node $\overline{\alpha}$, average repair bandwidth $\overline{\gamma}$, and $\ell$ compromised nodes.  The secrecy capacity of this system is upper bounded by
\begin{equation}\label{eq:secho}
C_s \leq C_s^{ho}(\overline{\alpha},\overline{\gamma},\ell).
\end{equation}
\end{theorem}
Equations \eqref{eq:secho} and \eqref{eq:secCap} imply the following upper bound stated in Th.~\ref{th:upsec}:
\begin{equation}
C_s\leq \sum_{i=\ell+1}^{k} \min\left\{ \overline{\alpha}, (d-i+1)\frac{\overline{\gamma}}{d}\right \}.
\end{equation}

Using Th.~\ref{eq:secgenup}, we easily deduce that symmetric repair is also optimal in terms of maximizing the secrecy capacity of a compromised DSS.
\begin{corollary}\label{cor:secsym}
Symmetric repair maximizes the secrecy capacity of a homogeneous system with a given budget on total repair bandwidth.
\end{corollary}

\section{Conclusion}\label{sec:conclusion}
We have studied distributed storage systems that are heterogeneous. Nodes in these systems can have different storage capacities and different repair bandwidths. We have focused on determining the information theoretic capacity of these systems, \emph{i.e.}, the maximum amount of information they can store, to achieve a required level of reliability (any $k$ out of the $n$ nodes should be able to give a stored file to a user). We have proved an upper bound on the capacity  that depends on the average resources available per node. Moreover, we have given an expression for the system capacity when we know all the nodes' parameters. This expression may be hard to compute, but we use it to derive additional upper and lower bounds that are easy to evaluate. We have also studied the case in which the system is compromised    by an eavesdropper, and have provided bounds on the system secrecy capacity under a perfect secrecy constraint. Our results
imply that symmetric repair maximizes the capacity of a homogeneous system, which justifies the repair  model used in the literature.
Problems that  remain open include finding an  efficient algorithm to compute the capacity of a heterogeneous distributed storage system, as well as efficient code constructions.
\begin{appendix}
\subsection{Functional vs. Exact Repair}\label{AppendixZ}
All of our results so far assumed   a functional repair model. However,  Theorems \ref{th:HCap}, \ref{th:upsec} and \ref{eq:secgenup} can be directly extended to the  exact repair case. For instance, Th.~\ref{th:HCap} becomes:
\begin{theorem}\label{th:HCapexact}
The capacity $C$ of a heterogeneous distributed storage system under exact repair, with node average capacity $\bar{\alpha}$ and average  repair bandwidth  $\bar{\gamma}$, is upper bounded by
\begin{equation}\label{eq:HCap}
C\leq C^{ho}_{exact}(\bar{\alpha},\bar{\gamma}),
\end{equation}
where $C^{ho}_{exact}(\bar{\alpha},\bar{\gamma})$ is the capacity of a homogeneous DSS under exact repair.
\end{theorem}

In the proofs of  Theorems \ref{th:HCap}, \ref{th:upsec} and \ref{eq:secgenup} we construct a new ``big" storage system using the original one as a building block. Hence, if we had  exact repair in the original system to start with,  we will have exact repair in the new ``big" system. The results can thus be   straightforwardly generalized to the case of exact repair. Moreover, under an exact repair constraint,  a homogeneous DSS with symmetric  repair maximizes capacity  under given average node storage and repair bandwidth budgets.

The other results, namely Theorems \ref{th:upnlw}, \ref{th:upnlw2}, and \ref{th:cap},   are proved using the analysis of the information flow graph. Therefore, It is not clear if there is an  obvious extension of these results  to the case of exact repair.

\subsection{Proof of Theorem~\ref{th:HCap}}\label{AppendixA}

We prove Th.~\ref{th:HCap} by making formal the argument of the example in Section~\ref{sec:ToyEx}. We start by describing the operation of adding, or combining, together multiple storage systems having  same number of nodes. Let  $\mathcal{DSS}_1, \mathcal{DSS}_2$ be two storage systems with nodes $v_1^1,\dots,v_n^1$ and $v_1^2,\dots,v_n^2$, respectively. The new system that we refer to as $\mathcal{DSS}$ obtained by combining $\mathcal{DSS}_1$ and $\mathcal{DSS}_2$ is  comprised of $n$ nodes, say $u_1,\dots,u_n$. Node $u_i$ has storage capacity $\alpha_i=\alpha_i^1+\alpha_i^2$ (superscript $j, j=1,2,$ denotes a parameter of system $S_j$). Moreover,  when node $u_j$ fails in $DSS$, the new node downloads $\beta_{ijS}=\beta_{ijS}^1+\beta_{ijS}^2$ amount of information from helper node $u_i$ (recall that $S$ is the set of indices of the $d$ helper nodes). We write $\mathcal{DSS}=\mathcal{DSS}_1+\mathcal{DSS}_2$.

Now, let $\mathcal{DSS}$ be the given heterogeneous system for which we wish to compute its capacity $C$. For each permutation $\sigma:\{1,\dots,n\}\rightarrow\{1,\dots,n\}$, we denote by $\mathcal{DSS}_\sigma$ the storage system with nodes $v_1^\sigma,\dots,v_n^\sigma$ such that $v_i^\sigma=v_{\sigma(i)}$. Let $\mathcal{P}_n$ denote the set of all $n!$ permutations on the set $\{1,\dots,n\}$. We define a new ``big" system  by $$\mathcal{DSS}_b=\sum_{\sigma\in\mathcal{P}_n}\mathcal{DSS}_\sigma.$$
The new system $\mathcal{DSS}_b$ is homogeneous with symmetric repair where the  storage capacity per node $\alpha_b$ is given by
$$
\alpha_b=(n-1)!\sum_{i=1}^{n}\alpha_i=n!\overline{\alpha},
$$
and the repair bandwidth per helper $\beta_b$ is given by
\begin{equation}
\begin{split}
\beta_b=& (n-d-1)!(d-1)!\sum_{j=1}^{n}\sum_{\substack{
           i=1 \\
           i \neq j}}^{n}\sum_{\substack{
           S \\
           i\in S \\
           j\notin S \\
           |S|=d}} \beta_{ijS} \\
      =& (n-d-1)!(d-1)!\sum_{j=1}^{n}\binom{n-1}{d}\bar{\gamma}_j \\
      =& \frac{(n-1)!}{d}\sum_{j=1}^{n}\bar{\gamma}_j=n!\frac{\bar{\gamma}}{d}.
\end{split}
\end{equation}

Therefore, the capacity $C_b$ of $\mathcal{DSS}_b$ as given by \eqref{eq:AlexCap} is

\begin{equation}\label{eq:big}
C_b=n!\sum_{i=1}^{k}\min\left\{\bar{\alpha},(d-i+1)\frac{\bar{\gamma}}{d}\right\}.
\end{equation}

Any scheme achieving the capacity $C$ of the original system can be naturally extended to store a file of size $n!C$ in $\mathcal{DSS}_b$ (see Fig.~\ref{fig:DSS}). Therefore, $C_b\geq n! C$. This inequality combined with \eqref{eq:big} gives the result of the Th.~\ref{th:HCap}.

\subsection{Proof of Theorem~\ref{th:cap} (sketch)}\label{sec:A2}

We use the definition of the flow graph in \cite{DGWWR07} to represent the DSS. The flow graph is a multicast network in which the multiple destinations correspond to the users requesting files from the DSS by contacting any $k$ out of the $n$ nodes. Therefore, the capacity of the DSS is the capacity of this multicast network which is equal to the minimum value of the  min-cuts to the users, by the fundamental theorem of network coding.  Note that  in the flow graph, a storage node $v_i$ is represented by two vertices $x_{in}^i$ and $x_{out}^i$ connected by an edge of capacity $\alpha_i$ (see Fig.~\ref{fig:cut}).

Let $C$ be the capacity of the DSS and  define $F$ to be
$$
F\triangleq \min_{\substack{
(f_1, \dots, f_k) \\
f_i \neq f_j \text{ for } i \neq j }} \sum_{i=1}^{k} \min \left( \alpha_{f_i} , \min_{\substack{
|S_i|=d+1-i \\
\{ f_1 , \dots, f_i \} \cap S_i = \emptyset }} \beta_{S_i} \right).
$$
We want to show that $C=F$.

Let $(f_1, \dots, f_k)$ be fixed and consider the successive failures and repairs of nodes $v_{f_1},\dots,v_{f_n}$ as seen in Fig.~\ref{fig:cut}.
Suppose node  $v_{f_1}$ is repaired by  contacting the helper nodes that minimize the sum $\beta_{S_1}$ with $|S_1|=d$ and $\{ f_1 \} \cap S_1 = \emptyset$, and node $v_{f_2}$ is repaired by contacting node $v_{f_1}$ and the $d-1$ helper  nodes that minimize the sum $\beta_{S_2}$ with $|S_2|=d-1$ and $\{ f_1 , f_2 \} \cap S_2 = \emptyset$. We continue in this fashion and finish with node $v_{f_k}$ being repaired by contacting nodes $v_{f_1},\dots,v_{f_{k-1}}$ and the $d-k+1$ helper nodes that minimize $\beta_{S_k}$ with $|S_k|=d+1-k$ and $\{ f_1 , \dots, f_k \} \cap S_k = \emptyset$. Now consider a user contacting nodes $v_{f_1},\dots,v_{f_n}$ there is a cut to the user of value $F$. By the max-flow min-cut theorem, we get $C\leq F$.

To prove the other direction, consider a user in the system and
 let $E$ denote the edges in the min-cut that separates  this user from the source in the flow graph. Also, let $V$ be the set of vertices in the flow graph that have a path to the user. Since the flow graph is acyclic, we have a topological ordering of the vertices  in $V$, which means that they can be indexed such that an edge from $v_i$ to $v_j$ implies $i<j$.

Let  $x_{out}^{1}$ be the first ``out-node" in $V$ (with respect to the ordering). If $x_{in}^{1} \notin V$, then $x_{in}^{1}x_{out}^{1} \in E$. On the other hand, if $x_{in}^{1} \in V$, then the set of incoming edges $S_1$, $|S_1|=d$, of $x_{in}^{1}$ must be in $E$.

Now similarly let $x_{out}^{2}$ be the second ``out-node" in $V$ with respect to the ordering. If $x_{in}^{2} \notin V$, then $x_{in}^{1}x_{out}^{2} \in E$. {If $x_{in}^{2} \in V$, then the set $S_2, |S_2| \geq d-1$, of edges incoming to  $x_{in}^{2}$, not including a possible edge from $x_{out}^{1}$,  must be in $E$.} All $k$ nodes adjacent to the user must be in $V$ so continuing in the same fashion gives that the min-cut is at least
$$
\sum_{i=1}^{k} \min( \alpha_{f_i} , \beta_{S_i}),
$$
where $f_i \neq f_j \text{ for } i \neq j$, $|S_i| = d+1-i$, and $\{ f_1 , \dots, f_i \} \cap S_i = \emptyset$. Hence $C \geq F$.

\end{appendix}

\end{document}